\documentclass[cup5b]{caps}
\usepackage{graphicx}
\usepackage{amssymb}
\usepackage{ociwsymp3e}  

\HeadText{T. E. Jeltema et al.}  

\def\plotone#1{\centering \leavevmode
\includegraphics[width=.95\columnwidth]{#1}}

\def\plotone#1{\centering \leavevmode
\includegraphics[width=.95\columnwidth]{#1}}

\begin{document}

\pagenumbering{arabic}

\author[]{T. E. JELTEMA$^{1}$, C. R. CANIZARES$^{1}$, M. W. BAUTZ$^{1}$, and D. A. BUOTE$^{2}$
\\
(1) Massachusetts Institute of Technology, Boston, MA, USA\\
(2) University of California, Irvine, Irvine, CA, USA}

\chapter{The Evolution of Cluster Substructure}

\begin{abstract}

Using Chandra archival data, we have begun a study to quantify the evolution of cluster morphology with redshift.  To quantify cluster morphology, we use the power ratio method developed by Buote and Tsai (1995).  Power ratios are constructed from moments of the two-dimensional gravitational potential and are, therefore, related to a cluster's dynamical state.  Our sample will include around 50 clusters from the Chandra archive with redshifts between 0.11 and 1.1.  These clusters were selected from two fairly complete flux-limited X-ray surveys (the ROSAT Bright Cluster Sample and the Einstein Medium Sensitivity Survey), and additional high-redshift clusters were selected from recent ROSAT flux-limited surveys.  Here we present preliminary results from the first 15 clusters in this sample.  Of these, eight have redshifts below 0.5, and seven have redshifts above 0.5.

\end{abstract}

\section{Introduction and Sample Selection}

Substructure or a disturbed cluster morphology indicates that a cluster is dynamically young (i.e. it will take some time for it to reach a relaxed state).  The amount of substructure in clusters in the present epoch and how quickly it evolves with redshift depend on the underlying cosmology.  In low density universes, clusters form earlier and will be on average more relaxed in the present epoch.  Clusters at high redshift, closer to the epoch of cluster formation, should be on average dynamically younger and show more substructure.  In addition to placing constraints on the mean matter density of the universe, the evolution of cluster morphology is important to the understanding of many cluster properties including mass, gas mass fraction, lensing properties, and galaxy morphology and evolution.

Several studies have been done to quantify substructure in clusters at low redshift (e.g., Jones \& Forman 1992; Mohr et al. 1995; Buote \& Tsai 1996).  However, it is only with recent X-ray and optical surveys that we are beginning to find tens of clusters with z $>$ 0.8, and it is becoming possible to study the evolution of substructure.  Using the power ratio method (Buote \& Tsai 1995), we are studying structure in a sample of approximately 50 clusters observed with the Chandra X-ray Observatory.  As a first cut, our sample includes only clusters with a redshift above 0.1 so that a reasonable area of each cluster will fit on a Chandra CCD.  In order to have a reasonably unbiased sample, clusters were selected from the BCS (Ebeling et al. 1998) and EMSS (Gioia \& Luppino 1994) surveys.  They were also required to have a luminosity greater than $5 \times 10^{44}$ ergs s$^{-1}$, as listed in those catalogs.  Additional high-redshift clusters were selected from recent ROSAT flux-limited surveys (Rosati et al. 1998; Perlman et al. 2002; Voges et al. 2001; Vikhlinin et al. 1998).  This led to a sample of approximately 50 clusters with redshifts between 0.11 and 1.1.  Here we present the results from 15 of these clusters.  Eight of these have redshifts below 0.5 with an average redshift of 0.25; the other seven have redshifts above 0.5 and an average redshift of 0.71.

\section{Power Ratios}

Power ratios are constructed from moments of the two-dimensional gravitational potential and are capable of distinguishing a large range of cluster morphologies.  Specifically, one calculates the squares of the moments over a circle of radius, R, centered on the cluster's center of mass.  Here we use an aperture radius of 0.5 Mpc for all clusters.  To normalize, one then takes the ratio of the {\itshape m\/}th multipole term to the monopole term.

The {\itshape m\/}th power ratio is given by
\begin{equation}
{P_m\over P_0} \equiv
{ \langle(\Psi^{\rm int}_m)^2\rangle\over \langle(\Psi^{\rm int}_0)^2\rangle}, \nonumber
\end{equation}
where $\Psi_m^{\rm int}$ is the $m$th multipole of the 2D gravitational potential and $\langle\cdots\rangle$ represents the azimuthal average around the circle.  In a little more detail, the multipole expansion of the two-dimensional gravitational potential is
\begin{equation}
\Psi(R,\phi) = -2Ga_0\ln\left({1 \over R}\right) -2G
\sum^{\infty}_{m=1} {1\over m R^m}\left(a_m\cos m\phi + b_m\sin
m\phi\right). \label{eqn.multipole}
\end{equation}
The moments $a_m$ and $b_m$ are
\begin{eqnarray}
a_m(R) & = & \int_{R^{\prime}\le R} \Sigma(\vec x^{\prime})
\left(R^{\prime}\right)^m \cos m\phi^{\prime} d^2x^{\prime}, \nonumber \\
b_m(R) & = & \int_{R^{\prime}\le R} \Sigma(\vec x^{\prime})
\left(R^{\prime}\right)^m \sin m\phi^{\prime} d^2x^{\prime}, \nonumber
\end{eqnarray}
where $\vec x^{\prime} = (R^{\prime},\phi^{\prime})$, and $\Sigma$ is the surface mass density (or surface brightness for X-ray observations).  The powers are then given by
\begin{equation}
P_0=\left[a_0\ln\left(R\right)\right]^2 \nonumber
\end{equation}
\begin{equation}
P_m={1\over 2m^2 R^{2m}}\left( a^2_m + b^2_m\right). \nonumber
\end{equation}

In the case of X-ray studies, X-ray surface brightness replaces surface mass density in the calculation of power ratios.  X-ray surface brightness is proportional to gas density squared and generally shows the same qualitative structure as the projected mass density, allowing a similar quantitative classification of clusters.

\section{Preliminary Results}

For each cluster in our initial sample of 15, we calculated P$_2$/P$_0$, P$_3$/P$_0$, and P$_4$/P$_0$ centered on the cluster centroid (where P$_1$ vanishes).  For all three power ratios, the high-redshift clusters have higher average power ratios then the low-redshift clusters, indicating more structure.  However, we would like to know if this result is statistically significant.  A Mann-Whitney rank-sum test shows that the P$_3$/P$_0$ ratios for the high-redshift clusters are on average larger than those for the low-redshift clusters at 99\% significance.  The P$_4$/P$_0$ ratios are also higher for the high-redshift clusters at 90\% significance.  Also, a Kolmogorov-Smirnov (KS) test of the P$_3$/P$_0$ ratios for the high-redshift sample versus the low-redshift sample gives a probability of 6\% that they drawn from the same distribution.  However, a KS test of the P$_2$/P$_0$ and P$_4$/P$_0$ ratios for the two samples does not show a significant difference in the distributions, and a Mann-Whitney rank-sum test of the P$_2$/P$_0$ ratios also shows no significant difference.  One possible reason that P$_3$/P$_0$ is better at distinguishing the high-redshift clusters from the low-redshift ones is that it is not sensitive to ellipticity: a single elliptical cluster will only have even multipoles.  Large odd multipoles unambiguously indicate asymmetry (substructure) in a cluster.

In Figure 1.1, we plot P$_2$/P$_0$ versus P$_3$/P$_0$ for the preliminary sample of 15 clusters.  High-redshift clusters (z$>$0.5) are plotted with diamonds.  Low-redshift clusters are plotted with asterisks.  As an example of where various cluster morphologies fall on this plot, we show adaptively smoothed Chandra images of three high and three low-redshift clusters with their power ratios indicated.  In the top-right corner, there is a double cluster (CL0152) and a very complicated looking cluster (V1121).  In the bottom-left corner is RXJ0439+05, a very relaxed looking cluster.  In between these are an elliptical cluster and a couple of clusters with smaller scale substructure.  It is clear from this plot that the high-redshift clusters do generally have higher P$_3$/P$_0$ than the low-redshift clusters.  For comparison, in Figure 1.2 we show P$_2$/P$_0$ versus P$_3$/P$_0$ for a sample of low-redshift clusters observed with ROSAT PSPC (Buote \& Tsai 1996).


\begin{figure}
   \centering
   \caption{Power ratios (P$_2$/P$_0$ versus P$_3$/P$_0$) computed in a 0.5 Mpc aperture for the first 15 clusters in our sample.  High-redshift clusters (z$>$0.5) are plotted with diamonds.  Low-redshift clusters are plotted with asterisks.  Also shown are adaptively smoothed Chandra images of three high and three low-redshift clusters.  The power ratios of these clusters are indicated.}
   \label{sample-figure}
\end{figure}

\begin{figure}
\plotone{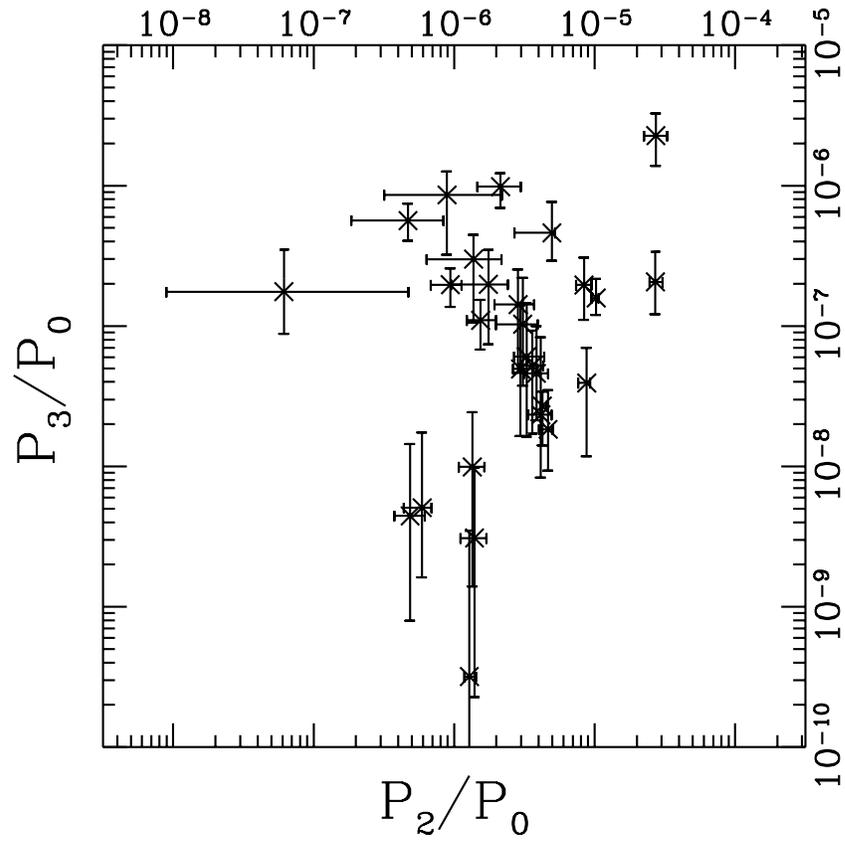}
\caption{Power ratios (P$_2$/P$_0$ versus P$_3$/P$_0$) computed in a 0.5 Mpc aperture for a sample of clusters observed with ROSAT PSPC (from Buote \& Tsai 1996).  All of these clusters have z$\leq$0.2.}
\end{figure}

At this point, we have only analyzed 15 out of 50 clusters in our sample, so we cannot make any definite conclusions.  However, this preliminary sample suggests that the amount of structure in clusters increases with redshift, a result which we hope to confirm and quantify with the full cluster sample.  We also plan to compare the sample to numerically simulated clusters.

\end{document}